\documentclass{pasa}%

\usepackage{graphicx}
\title[MyCn~18]{SALT HRS discovery of the binary nucleus of the Etched Hourglass Nebula MyCn~18\thanks{Based on observations made with the Southern African Large Telescope (SALT) under programmes 2016-2-SCI-034 and 2017-1-MLT-010.}}

\author[Miszalski et al.]{Brent Miszalski$^{1,2}$\thanks{E-mail: brent@saao.ac.za}, Rajeev Manick$^{3}$, Joanna Miko{\l}ajewska$^{4}$, Hans Van Winckel$^{3}$ and Krystian I{\l}kiewicz$^{4}$
\affil{$^1$South African Astronomical Observatory, PO Box 9, Observatory, 7935, South Africa}%
\affil{$^2$Southern African Large Telescope Foundation, PO Box 9, Observatory, 7935, South Africa}
\affil{$^3$Instituut voor Sterrenkunde, KU Leuven, Celestijnenlaan 200D bus 2401, B-3001 Leuven, Belgium}
\affil{$^4$Nicolaus Copernicus Astronomical Center, Polish Academy of Sciences, Bartycka 18, PL-00716 Warsaw, Poland}
}%

\jid{PASA}
\doi{10.1017/pas.\the\year.xxx}
\jyear{\the\year}

\usepackage{aas_macros}

\begin{document}

\begin{frontmatter}
\maketitle

\begin{abstract}
   The shaping of various morphological features of planetary nebulae (PNe) is increasingly linked to the role of binary central stars. Identifying a binary within a PN offers a powerful tool with which to directly investigate the formation mechanisms behind these features. The Etched Hourglass Nebula, MyCn~18, is the archetype for several binary-linked morphological features, yet it has no identified binary nucleus. It has the fastest jets seen in a PN of 630 km s$^{-1}$, a central star position offset from the nebula centre, and a bipolar nebula with a very narrow waist. Here we report on the Southern African Large Telescope (SALT) High Resolution Spectrograph (HRS) detection of radial velocity variability in the nucleus of MyCn~18 with an orbital period of $18.15\pm0.04$ days and a semi-amplitude of $11.0\pm0.3$ km s$^{-1}$. Adopting an orbital inclination of $38\pm5$ deg and a primary mass of $0.6\pm0.1$ $M_\odot$ yields a secondary mass of $0.19\pm0.05$ $M_\odot$ corresponding to an M5V companion. The detached nature of the binary rules out a classical nova (CN) as the origin of the jets or the offset central star as hypothesised in the literature. Furthermore, scenarios that produce the offset central star during the AGB and that form narrow waist bipolar nebulae result in orbital separations 80--800 times larger than observed in MyCn~18. The inner hourglass and jets may have formed from part of the common envelope ejecta that remained bound to the binary system in a circumbinary disk, whereas the offset central star position may best be explained by proper motion. Detailed simulations of MyCn~18 are encouraged that are compatible with the binary nucleus to further investigate its complex formation history. 
\end{abstract}

\end{frontmatter}

\section{Introduction}
\label{sec:intro}
At least 1 in 5 planetary nebulae host a close binary central star with an orbital period of $\sim$1 day or less (Bond 2000; Miszalski et al. 2009). These binaries have passed through the poorly understood common envelope phase of binary stellar evolution (Ivanova et al. 2013). The binary fraction of planetary nebulae is expected to be considerably higher if binary interactions are a dominant formation channel (De Marco 2009). Binary central stars with intermediate orbital periods of several days to years may contribute another 23--27\% towards the binary fraction (Nie et al. 2012), but it is unclear whether the small number of discoveries made so far (Van Winckel et al. 2014; Jones et al. 2017; Miszalski et al. 2018) signify the existence of a substantial population of intermediate period binaries. We are systematically searching for this population (Miszalski et al. 2018) with the High Resolution Spectrograph (HRS, Bramall et al. 2010, 2012; Crause et al. 2014) on the Southern African Large Telescope (SALT, Buckley et al. 2006; O'Donoghue et al. 2006). Further details of the scientific motivation behind the survey can be found in Miszalski et al. (2018).

\begin{figure}
   \begin{center}
      \includegraphics[scale=0.30,bb=0 0 800 800]{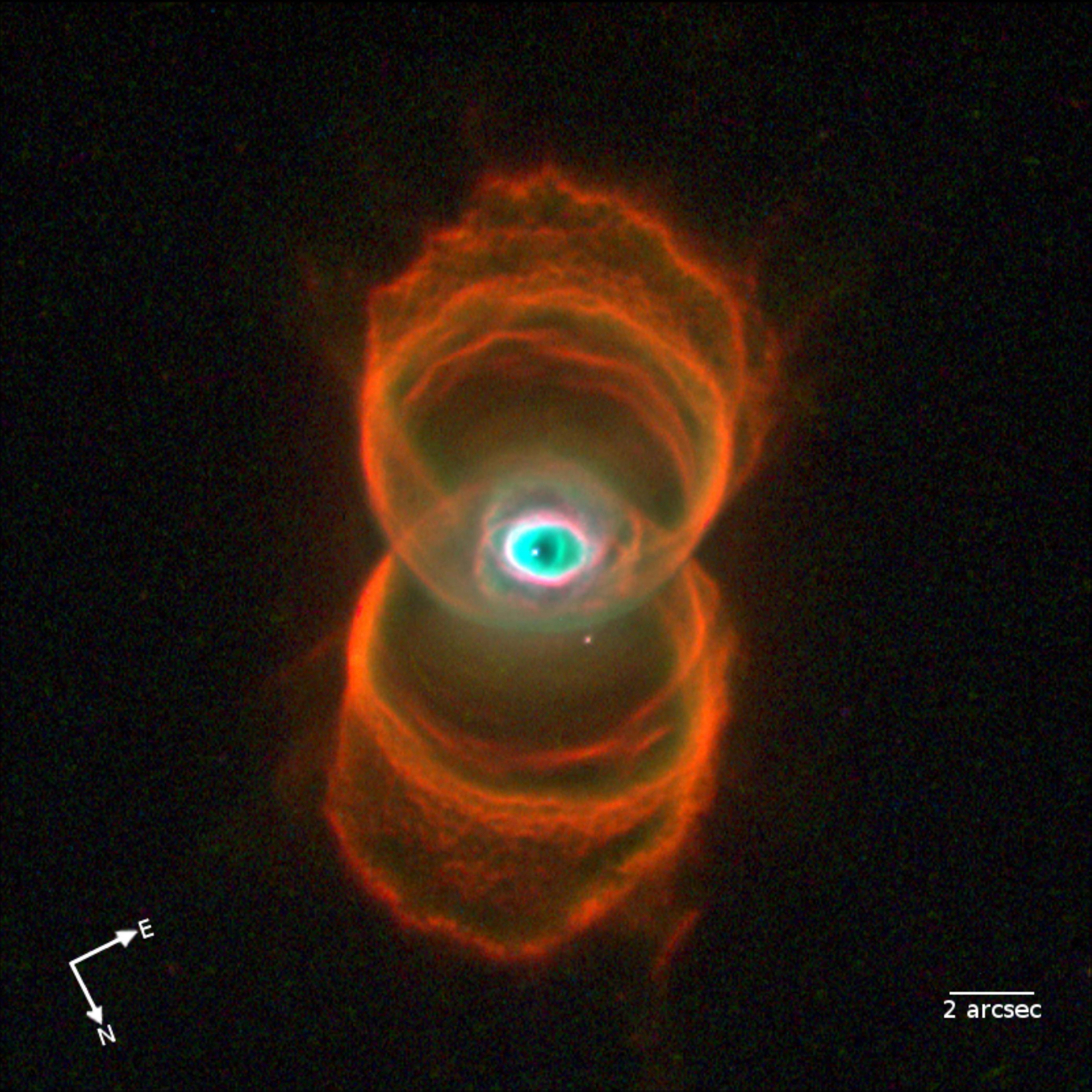}
   \end{center}
   \caption{\emph{Hubble Space Telescope} colour-composite image of MyCn~18 (Sahai et al. 1999) made from F658N (red), F656N (green) and F502N (blue) filters. Image credit: Raghvendra Sahai and John Trauger (JPL), the WFPC2 science team, and NASA.}
   \label{fig:img}
\end{figure}

Included in our survey is the young planetary nebula MyCn~18 (PN G307.5$-$04.9, Mayall \& Cannon 1940), also known as the Etched Hourglass Nebula. The bipolar appearance of MyCn~18 (Fig. \ref{fig:img}; Sahai et al. 1999) is the result of an eponymous hourglass-shaped nebula inclined 38 deg to the line of sight (Dayal et al. 2000; O'Connor et al. 2000; Clyne et al. 2014). MyCn~18 exhibits several features for which binary interactions are the preferred formation mechanism, yet no evidence has yet been found supporting the presence of a binary companion. These features include polar outflows or jets (Bryce et al. 1997; O'Connor et al. 2000), a central star position offset up to 0.2'' from the geometric centre and its bipolar morphology with a narrow waist (Sahai et al. 1999; Clyne et al. 2014). MyCn~18 has often been compared against bipolar nebulae around low and high mass evolved stars. It is particularly notable for its intriguing nested hourglass structure (Sahai et al. 1999; Clyne et al. 2014) that closely resembles the nebular remnant of SN 1987A (e.g. Burrows et al. 1995; Sugerman et al. 2015), the symbiotic nebula Hen~2-104 (Corradi \& Schwarz 1993; Corradi et al. 2001; Santander-Garc{\'{\i}}a et al. 2008; Clyne et al. 2015) and Hb~12 which may be a young PN or possibly a symbiotic nebula (Hsia et al. 2006; Kwok \& Hsia 2007; Vaytet et al. 2009; Clark et al. 2014). There are also strong similarities with bipolar and hourglass nebulae around several luminous blue variables (e.g. Garc{\'{\i}}a-Segura et al. 1999; Smith et al. 2007; Gvaramadze \& Menten 2012; Taylor et al. 2014). Based on these rich connections to several bipolar nebulae, resolving the unknown binary status of MyCn~18 is an important step towards further understanding how bipolar nebulae form. 

This paper is structured as follows. Section \ref{sec:obs} presents SALT HRS radial velocity monitoring observations of MyCn~18. We clarify the spectral class of the central star in Sect. \ref{sec:classification}, followed by a description of our radial velocity measurements in Sect. \ref{sec:rvmeas}. We demonstrate the significant periodic variability of these measurements in Sect. \ref{sec:params} that prove the binary nature of MyCn~18. Sect. \ref{sec:params} also derives the orbital parameters of the binary nucleus. We discuss our results in Sect. \ref{sec:discussion} in the context of formation scenarios concerning MyCn~18 and other post-CE PNe. We conclude in Sect. \ref{sec:conclusion}.

\section{SALT HRS Observations}
\label{sec:obs}

Multiple observations of the nucleus of MyCn~18 were taken with SALT HRS (Bramall et al. 2010, 2012; Crause et al. 2014), a dual-beam, fibre-fed \'echelle spectrograph enclosed in a vacuum tank within an insulated, temperature controlled enclosure in the spectrograph room of SALT (Buckley et al. 2006; O'Donoghue et al. 2006). The medium resolution (MR) mode of HRS was used to obtain spectra covering 3700--8900 \AA\ with resolving powers $R=\lambda/\Delta\lambda$ of 43000 and 40000 for the blue and red arms, respectively. During an observation a second fibre, separated at least 20 arcsec from the science fibre, simultaneously observes the sky spectrum. Both object and sky spectra are interleaved on the separate blue (2k x 4k) and red (4k x 4k) CCDs. Regular bias, ThAr arc lamp and quartz lamp flat field calibrations are taken as part of SALT operations. 

Table \ref{tab:log} presents a log of the 26 HRS MR observations of MyCn~18 in addition to the radial velocity measurements described in Sect. \ref{sec:rvmeas}. Basic processing of the data was performed by \textsc{pysalt} (Crawford et al. 2010) before a pipeline based on the \textsc{midas} packages \textsc{echelle} (Ballester 1992) and \textsc{feros} (Stahl et al. 1999) developed by A. Y. Kniazev (Kniazev et al. 2016) reduced the data and distributed the data products. The blue spectra which do not feature prominent sky emission lines were not sky subtracted and the red arm spectra were sky subtracted. The small angular size of MyCn~18 and the fibre fed design of HRS precluded the subtraction of nearby nebula emission. The order-merged spectra were converted to a logarithmic wavelength scale using \textsc{iraf} before heliocentric radial velocity corrections were added using the \textsc{velset} task of the \textsc{rvsao} package (Kurtz \& Mink 1998).

\begin{table*}
   \centering
   \caption{Observation log of SALT HRS spectra of MyCn~18 and radial velocity measurements. The Julian day represents the midpoint of each exposure and the radial velocity measurements were made from stellar N~III $\lambda$4634.14 \AA\ and nebular He~I $\lambda$4921.93 \AA\ (see Sect. \ref{sec:rvmeas}).}
   \begin{tabular}{lcccc}
      \hline
      Julian day & Exposure time & Orbital Phase & RV (N III) & RV (He I)\\
                 &         (s)   &               & (km s$^{-1}$) &  (km s$^{-1}$)\\
         \hline
    2457847.41564	&	3000&	0.26	&$-$61.67 $\pm$ 0.88	&$-$72.93 $\pm$ 0.06\\
    2457860.52551	&	3000&	0.99	&$-$73.97 $\pm$ 1.09	&$-$72.45 $\pm$ 0.06\\
    2457861.48087	&	3000&	0.04	&$-$70.40 $\pm$ 0.97	&$-$72.55 $\pm$ 0.05\\
    2457880.28752	&	2750&	0.07	&$-$70.29 $\pm$ 2.06	&$-$72.89 $\pm$ 0.07\\
    2457880.39974	&	2750&	0.08	&$-$68.18 $\pm$ 0.65	&$-$72.75 $\pm$ 0.05\\
    2457886.28001	&	2750&	0.40	&$-$71.17 $\pm$ 1.09	&$-$72.57 $\pm$ 0.08\\
    2457892.46150	&	2750&	0.74	&$-$81.86 $\pm$ 0.78	&$-$72.64 $\pm$ 0.06\\
    2457898.34154	&	2750&	0.07	&$-$65.30 $\pm$ 1.06	&$-$73.19 $\pm$ 0.07\\
    2457904.22737	&	2750&	0.39	&$-$63.75 $\pm$ 1.21	&$-$73.07 $\pm$ 0.07\\
    2457910.34825	&	3000&	0.73	&$-$86.15 $\pm$ 1.36	&$-$72.70 $\pm$ 0.06\\
    2457917.30923	&	2750&	0.11	&$-$65.67 $\pm$ 0.90	&$-$73.18 $\pm$ 0.06\\
    2457926.28171	&	2500&	0.60	&$-$82.39 $\pm$ 1.91	&$-$72.60 $\pm$ 0.06\\
    2457934.27854	&	2500&	0.04	&$-$66.21 $\pm$ 1.39	&$-$72.88 $\pm$ 0.06\\
    2457935.26325	&	3000&	0.10	&$-$64.56 $\pm$ 0.74	&$-$72.69 $\pm$ 0.06\\
    2457936.25671	&	3000&	0.15	&$-$62.04 $\pm$ 0.91	&$-$72.75 $\pm$ 0.06\\
    2457938.27615	&	2800&	0.26	&$-$59.02 $\pm$ 0.85	&$-$72.92 $\pm$ 0.07\\
    2457939.24394	&	2500&	0.32	&$-$60.49 $\pm$ 0.67	&$-$72.77 $\pm$ 0.06\\
    2457941.30262	&	2500&	0.43	&$-$65.85 $\pm$ 1.36	&$-$72.64 $\pm$ 0.08\\
    2457942.24390	&	2825&	0.48	&$-$73.50 $\pm$ 0.92	&$-$72.50 $\pm$ 0.07\\
    2457943.23967	&	2825&	0.54	&$-$68.95 $\pm$ 1.10	&$-$72.52 $\pm$ 0.07\\
    2457947.30832	&	2825&	0.76	&$-$86.20 $\pm$ 1.28	&$-$72.66 $\pm$ 0.06\\
    2457949.26982	&	2500&	0.87	&$-$76.58 $\pm$ 0.88	&$-$72.85 $\pm$ 0.06\\
    2457962.25088	&	2825&	0.58	&$-$78.31 $\pm$ 0.64	&$-$72.94 $\pm$ 0.06\\
    2457963.25458	&	2825&	0.64	&$-$76.18 $\pm$ 0.79	&$-$72.75 $\pm$ 0.06\\
    2457965.22577	&	2825&	0.75	&$-$81.75 $\pm$ 0.93	&$-$72.89 $\pm$ 0.06\\
    2457969.23510	&	2825&	0.97	&$-$74.17 $\pm$ 0.68	&$-$72.89 $\pm$ 0.07\\
         \hline
   \end{tabular}
   \label{tab:log}
\end{table*}

\section{Analysis}
\subsection{Spectral classification}
\label{sec:classification}
\begin{figure*}
   \begin{center}
      \includegraphics[scale=0.60,bb=0 0 720 720]{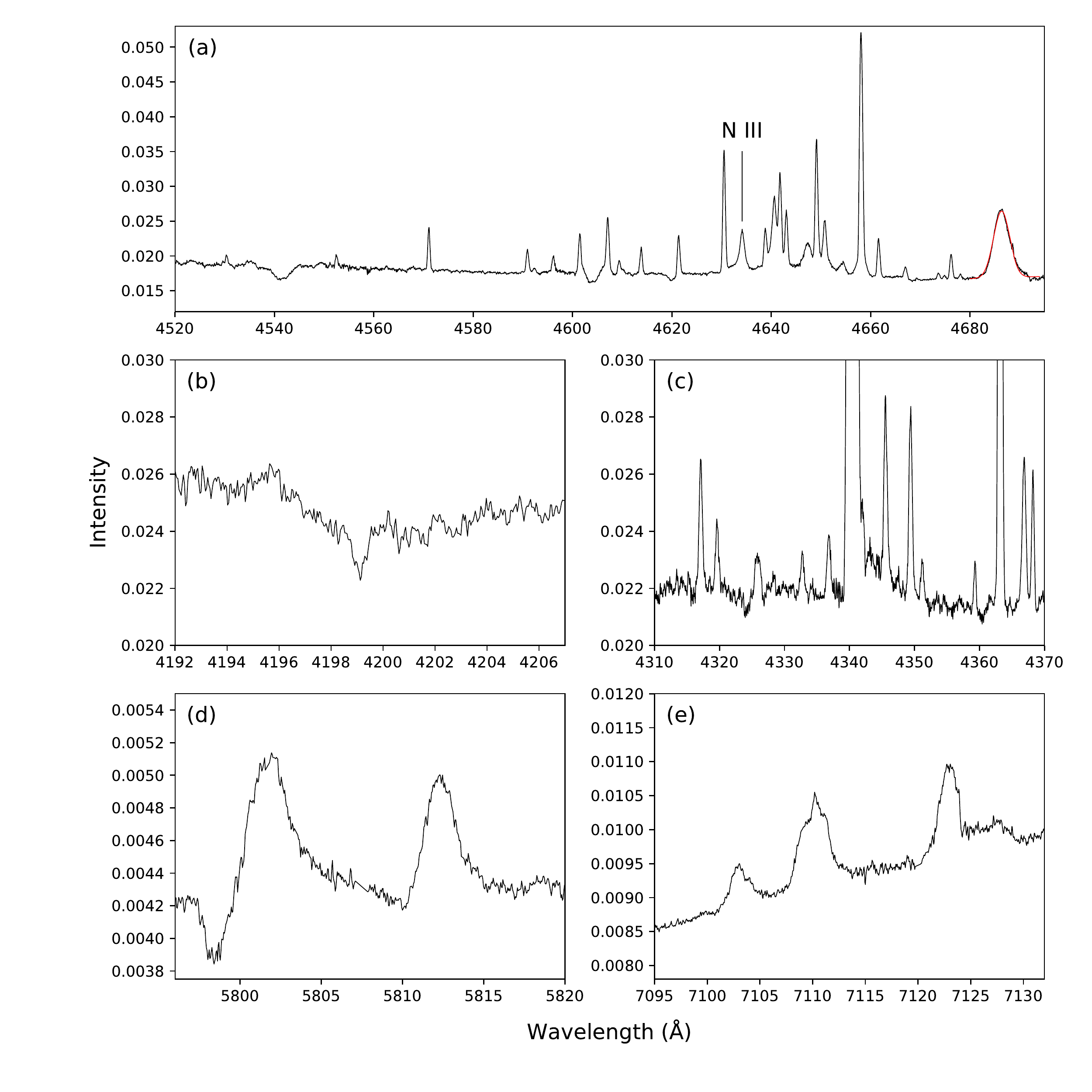}
   \end{center}
   \caption{Relevant portions of the average SALT HRS spectrum of the nucleus of MyCn~18. (a) He~II $\lambda$4540 \AA\ and $\lambda$4686 \AA, the latter overlaid with a Gaussian fit (red line, see text). Marked is the N~III $\lambda$4634 \AA\ line used to determine radial velocities. (b) He~II $\lambda$4200 \AA, (c) H$\gamma$ 4340 \AA, (d) C~IV 5801, 5812 \AA, (e) N~IV $\lambda$7103, 7109, 7111, 7123, 7127 and 7129 \AA.}
   \label{fig:cspn}
\end{figure*}

The spectral classification of the central star of MyCn~18 is unclear in the literature. An Of(H) classification was made based on an unpublished spectrum (M\'endez 1991) and more recently a rare H-deficient Of(C) classification was made by Lee et al. (2007) that requires a spectrum `dominated by strong C emissions' (M\'endez 1991). We created an average blue and red spectrum after shifting all HRS spectra to the rest frame of N~III $\lambda$4634 \AA\ using the measurements in Tab. \ref{tab:log} (see Sect. \ref{sec:rvmeas}), \textsc{iraf} and the \textsc{rvsao} package (Kurtz \& Mink 1998). Relevant portions are displayed in Fig. \ref{fig:cspn} which includes a Gaussian fit to He~II $\lambda$4686 \AA\ with a full-width at half-maximum (FWHM) of 3.83$\pm$0.03 \AA\ made using the \textsc{lmfit} package (Newville et al. 2016). The data do not support an Of(C) classification as the only prominent C emission features belong to C~IV $\lambda$4658, 5801, 5812 \AA. The C IV $\lambda$4658 \AA\ emission line is heavily contaminated in our spectrum by nebular [Fe III] emission (Tsamis et al. 2003).

The spectrum exhibits characteristics of both Of and Of-WR stars (M\'endez et al. 1990). According to the classification scheme for these stars (M\'endez et al. 1990), features supporting an Of classification include the narrow He~II $\lambda$4686 \AA\ emission (FWHM $<$ 4.0 \AA) and mildly blueshifted He~II $\lambda$4540 \AA\ absorption ($-12.5\pm0.7$ km s$^{-1}$), whereas supporting an Of-WR classification there is no absorption at H$\gamma$ (possibly with a red wing of emission) and blueshifted He~II $\lambda$4200 \AA\ more than $-50$ km s$^{-1}$ ($-52.1\pm0.7$ km s$^{-1}$). We favour an Of(H) classification based on the FWHM of He~II $\lambda$4686 \AA\ and strong similarities with He~II, C~IV and N~IV features in the Of(H) central star of M~2-29 (Miszalski et al. 2011a). The temperature of the central star is not well constrained. Gleizes et al. (1989) determined a H Zanstra temperature of 41 kK and a He~II Zanstra temperature of 51 kK based on low-resolution spectra. The stellar He~II emission may have been mistaken for nebular He~II emission in these spectra since the higher quality HRS spectra rule out any He~II nebular emission. Following the example of M~2-29 (Miszalski et al. 2011a), we therefore adopt instead $T_\mathrm{eff}=50\pm10$ kK as He~II nebular emission and He~I absorption lines are all absent from the HRS spectra.

\subsection{RV measurements}
\label{sec:rvmeas}
\begin{figure*}
   \begin{center}
      \includegraphics[scale=0.60,bb=0 0 720 936]{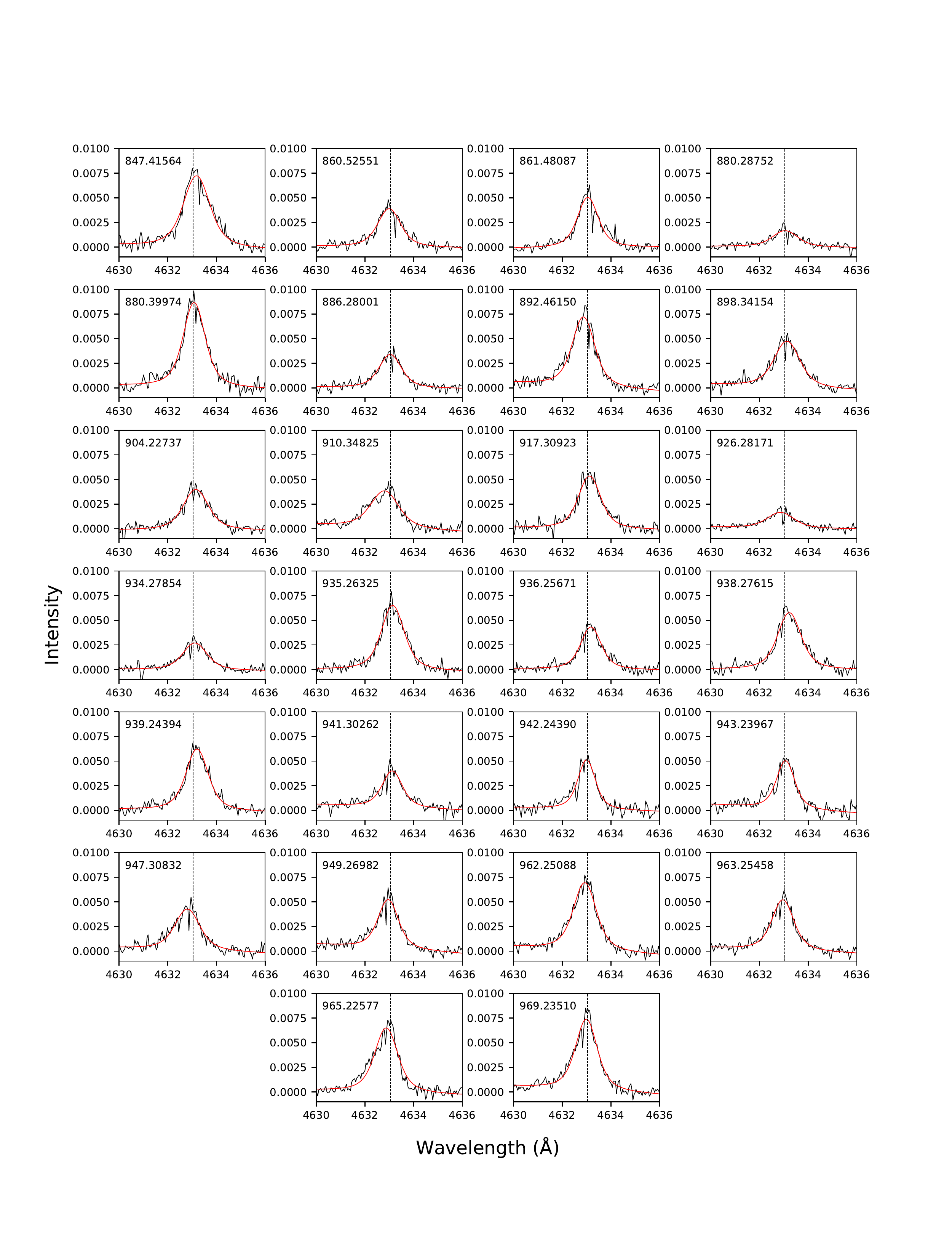}
   \end{center}
   \caption{The observed N~III $\lambda$4634.14 \AA\ profiles (black lines) were fit with a model composed of a Voigt function and a straight line (red lines). The dashed line represents the expected position of N~III at the systemic heliocentric radial velocity of the nebula of $-71$ km s$^{-1}$ (Clyne et al. 2014). Each panel is labelled with the Julian day of each spectrum minus 2457000 days.}
   \label{fig:multi}
\end{figure*}

The radial velocity measurements in Table \ref{tab:log} were determined from the stellar emission feature N~III $\lambda$4634.14 and the nebular emission line He~I $\lambda$4921.23. The \textsc{lmfit} package (Newville et al. 2016) was used to fit models consisting of a Voigt function and a straight line to N~III and a Gaussian function to He~I. The fits to N~III are displayed in Fig. \ref{fig:multi}. Uncertainties in each radial velocity measurement are 1$\sigma$ uncertainties determined from the standard error in the fit centroid provided by the fitting routine. The average nebular radial velocity of $-72.78\pm0.10$ km s$^{-1}$ agrees well with $-71$ km s$^{-1}$ measured from spatio-kinematic modeling (Clyne et al. 2014).

\subsection{Periodic variability and orbital parameters}
\label{sec:params}
\begin{figure*}
   \begin{center}
      \includegraphics[scale=0.45,bb=18 180 594 612]{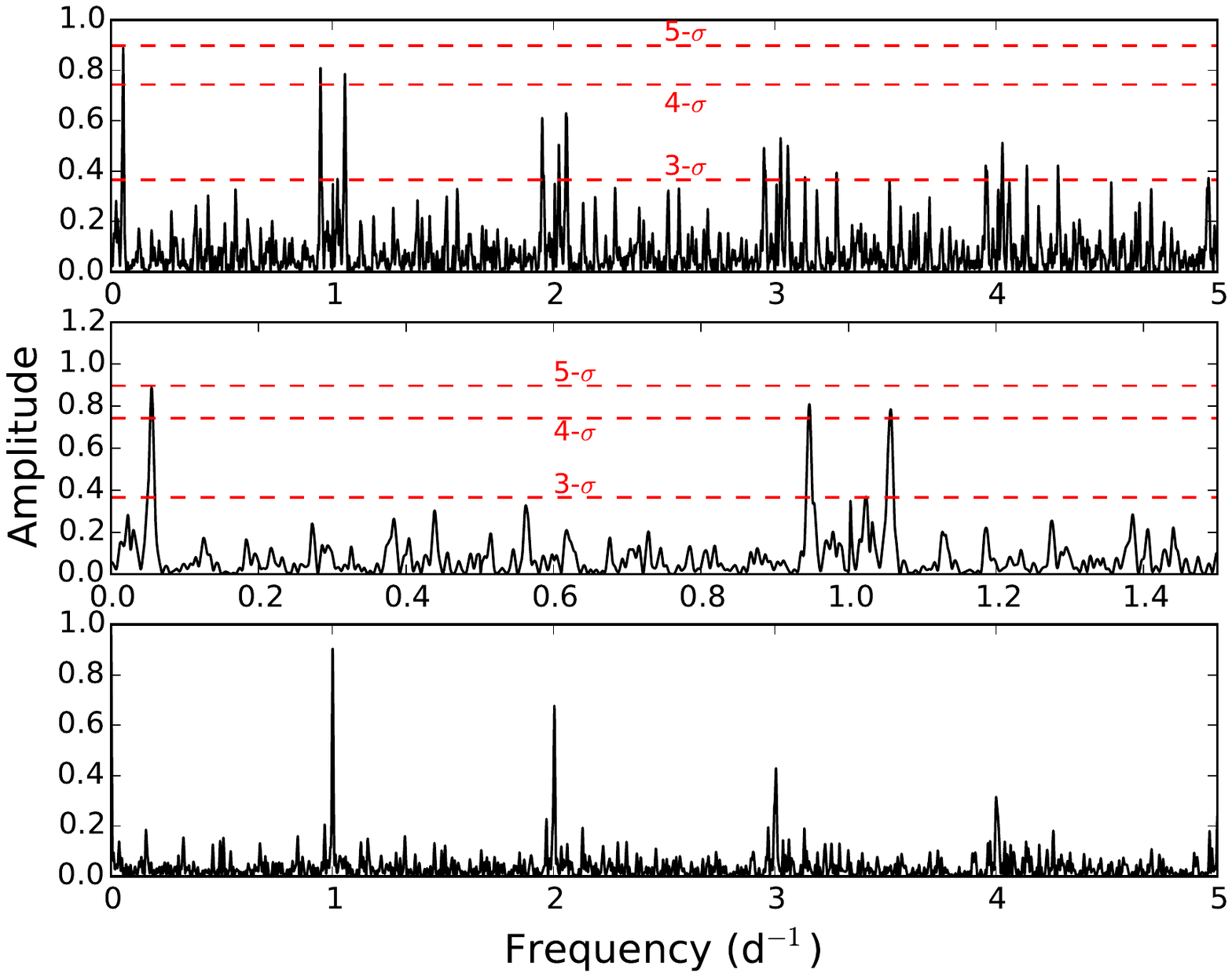}
      \includegraphics[scale=0.45,bb=18 180 594 612]{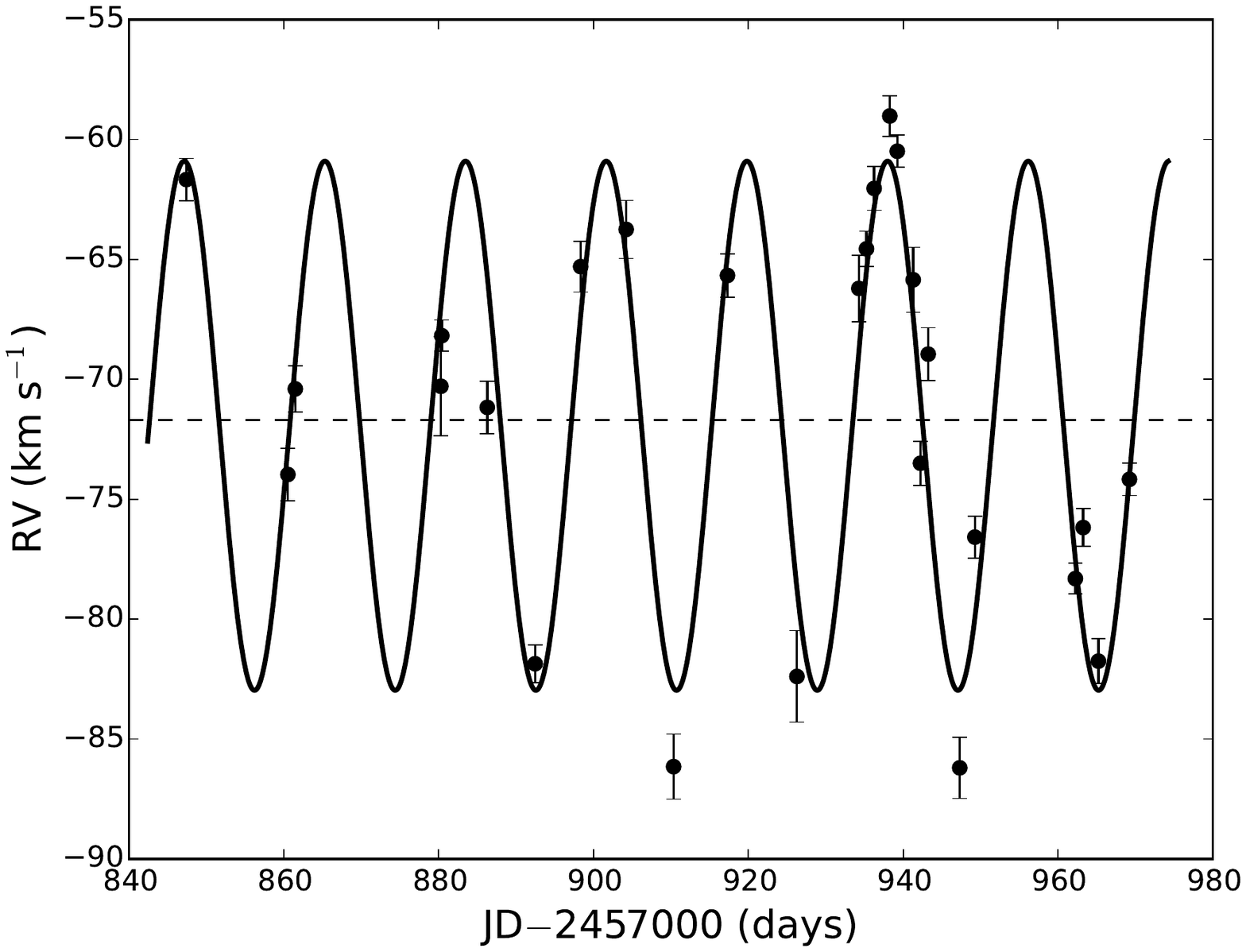}
      \includegraphics[scale=0.45,bb=18 180 594 612]{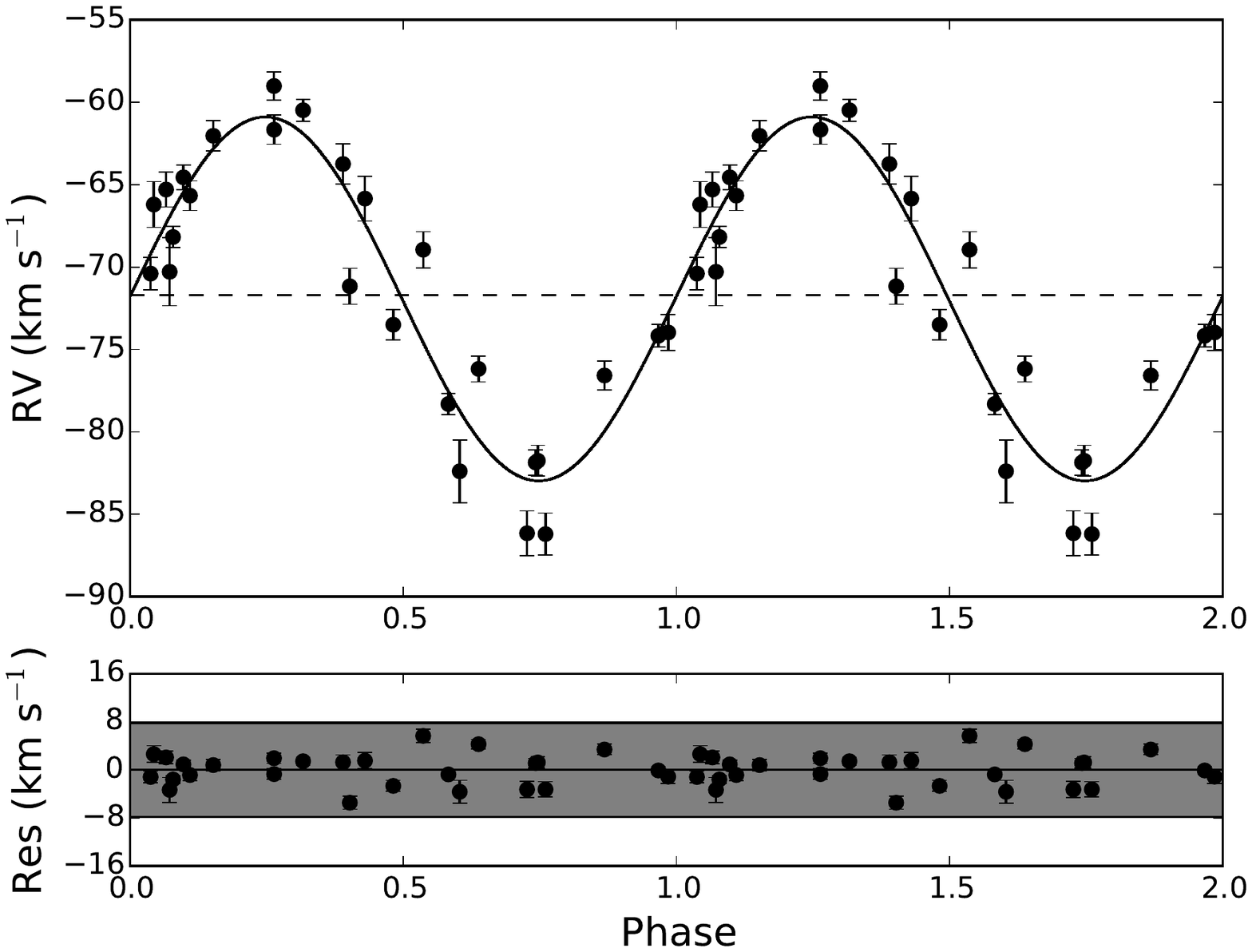}
   \end{center}
   \caption{The binary nature of MyCn 18 revealed by SALT HRS. \emph{(Top panel)} The Lomb-Scargle periodogram of radial velocity measurements (top two segments). The strongest peak at a $5\sigma$ significance level corresponds to the 18.15 d orbital period. The $1-f$ and $1+f$ aliases are also visible and the lower segment shows the window function. \emph{(Middle and Bottom panels)} Radial velocity measurements displayed in time (middle) and folded with the orbital period (bottom). The solid lines represent the Keplerian orbit fit and the shaded region indicates the residuals are within $3\sigma$ of the fit where  $\sigma=2.61$ km s$^{-1}$.}
   \label{fig:rvs}
\end{figure*}

The SALT HRS radial velocity measurements of the nucleus of MyCn~18 exhibit significant variability compared to the stationary nebular emission (Tab. \ref{tab:log}). Period analysis of the radial velocities using a Lomb-Scargle periodogram (Press et al. 1992) reveals the strongest peak to have a period of 18.15 d, significant at the 5$\sigma$ level, while other peaks correspond to the $1-f$ and $1+f$ aliases of this peak (Fig. \ref{fig:rvs}). The 18.15 d period was used as the basis for a Keplerian orbit model that was built using a least-squares minimization method applied to the phase-folded data. Figure \ref{fig:rvs} displays the Keplerian orbit fit overlaid on the radial velocity measurements in time and folded with the orbital period. The observed sinusoidal radial velocity shifts prove the binary nature of MyCn~18.

Table \ref{tab:orb} presents the orbital parameters of the Keplerian orbit fit determined using Monte Carlo simulations in which the eccentricity was fixed to be zero (for details see Miszalski et al. 2018). The systemic velocity of the orbit ($V_\mathrm{hel}=-71.70\pm0.24$ km s$^{-1}$; $V_\mathrm{LSR}=-65.20\pm0.24$ km s$^{-1}$) agrees well with the nebular radial velocity of $-71$ km s$^{-1}$ (Clyne et al. 2014). This suggests the N~III emission line from the Of(H)-type primary is not significantly disturbed by the stellar wind and traces the motion of the primary. 

\begin{table*}
   \centering
   \caption{Orbital parameters of the binary nucleus of MyCn~18.}
   \begin{tabular}{ll}
      \hline
Orbital period $P$ (d)                           &	$18.15\pm0.04$ \\
Eccentricity $e$	                               & 0 (fixed)       \\
Radial velocity semi-amplitude $K$ (km s$^{-1}$) &	$11.0\pm0.3$    \\
Systemic heliocentric velocity $\gamma$ (km s$^{-1}$)	       & $-71.70\pm0.24$ \\
Separation of primary from centre of mass $a_1\sin i$ (AU)	& $0.0183\pm0.0005$\\
Mass function $f(m)$ ($M_\odot$)                 &	$0.0025\pm0.0002$ \\
Epoch at radial velocity minimum $T_0$ (d)	       & $2457965.257\pm0.043$ \\
Root-mean-square residuals of Keplerian fit (km s$^{-1}$) & 	2.61 \\
Inclination $i$ (degrees)                        &	$38\pm5$ \\
Mass of primary $M_1$ ($M_\odot$)                &	$0.60\pm0.10$ \\
Mass of secondary $M_2$ ($M_\odot$)              & $0.19\pm0.05$ \\
      \hline
   \end{tabular}
   \label{tab:orb}
\end{table*}

The estimation of the secondary mass requires some assumptions concerning the orbital inclination $i$ and the primary mass $M_1$. Spatio-kinematic studies of post-CE PNe have demonstrated that the orbital inclination of the binary central star always matches the inclination determined from analysis of the nebula (Hillwig et al. 2016). The inclination of MyCn~18 is well determined to be 38 deg by spatio-kinematic studies (O'Connor et al. 2000; Clyne et al. 2014), although no estimate of the uncertainty was provided by these studies. We have therefore adopted a nominal error of 5 deg in the orbital inclination.

The primary mass $M_1$ is more difficult to constrain with the information available in the literature. Masses may be estimated via comparison with post-AGB evolutionary tracks if the temperature and surface gravity or luminosity are known. Determining the surface gravity is complicated by the strong stellar wind which has likely contaminated the absorption line profiles. We therefore adopt an alternative distance-dependent approach that estimates the luminosity of the central star following Sect. 9.4.5 of Frew (2008). We adopt a weighted mean distance of $3092\pm507$ pc, which incorporates distance estimates of $3342\pm668$ pc (Stanghellini \& Haywood 2010) and $2750\pm780$ pc (Frew et al. 2016), along with an apparent magnitude $m_V=14.9$ mag (Sahai et al. 1999), $T_\mathrm{eff}=50\pm10$ kK and a reddening of $A_V=2.63$ mag (Tsamis et al. 2003). We determine using equation 9.13 of Frew (2008) a bolometric correction of $-4.48\pm0.66$ mag and an absolute bolometric magnitude of $-4.67\pm0.77$, where the error in the latter includes errors in the bolometric correction and absolute magnitude added in quadrature which are themselves dominated by the uncertainties in the temperature and the distance, respectively. This results in $\log L/L_\odot=3.76\pm0.31$ which is typical for young central stars of PNe still on the horizontal part of post-AGB evolutionary tracks. Considering the uncertainties in this distance-dependent method, we adopt $M_1=0.6\pm0.1$ $M_\odot$ after making comparisons with post-AGB evolutionary tracks (Miller Bertolami et al. 2016). It is unlikely that the mass is above this range as the nebular chemical abundances of MyCn~18 are not consistent with a massive AGB progenitor that has a core mass above $\sim$0.8 $M_\odot$ (Garc{\'{\i}}a-Hern{\'a}ndez et al. 2016). This is dependent, however, on whether single star AGB models are applicable to post-CE PNe, which is yet to be determined. 

The radial velocity semi-amplitude of $K=11.0\pm0.3$ km s$^{-1}$, $M_1=0.6\pm0.1$ $M_\odot$ and $i=38\pm5$ deg therefore gives a secondary mass of $M_2=0.19\pm0.05$ $M_\odot$. According to the Teff-radius-mass relation for dwarfs with Z=0.014 (Bressan et al. 2012; Chen et al. 2014), the range of masses and temperatures give an M5V companion with an uncertainty of one spectral class (Rajpurohit et al. 2013). The current orbital separation is 0.124 au or 26.6 $R_\odot$ and the M5V companion has a Roche lobe radius of $\sim$7.5 $R_\odot$. Even if the radius were inflated by $\sim$2--3 times as observed in some post-CE binaries (Af{\c s}ar \& \.{I}bano{\v g}lu 2008), mass transfer via Roche lobe overflow would not be possible. With the aforementioned distance and reddening the M5V companion would be very faint at $V>21$ mag (Covey et al. 2007), explaining the lack of any secondary features in our spectra. A periodic photometric signal caused by the irradiated atmosphere of the companion may be present with an amplitude of $\sim$0.1 mag (De Marco et al. 2008). No photometric monitoring observations have been reported in the literature, but its detection would be complicated by the bright inner nebula.

\section{Discussion}
\label{sec:discussion}
\subsection{A classical nova could not have launched the jets of MyCn~18}
\label{sec:nova}
The system of collimated outflows or jets of MyCn~18 are remarkable as the fastest observed amongst PNe with deprojected velocities of up to 630 km s$^{-1}$ (Bryce et al. 1997; O'Connor et al. 2000; Clyne et al. 2014). The jets were ejected $\sim$1050 years after the main nebula (Clyne et al. 2014). Sahai et al. (1999) estimated the mass of the knots making up the jet system of MyCn~18 to be $\sim$10$^{-5}$ $M_\odot$. O'Connor et al. (2000) suggested that since classical nova (CN) ejecta have similar masses and kinematics, it may be that a CN was responsible for producing the jet system of MyCn~18. More recently, Soker \& Kashi (2012) supported the assessment of O'Connor et al. (2000) and further commented that the kinetic energy of the jets was too low to have been formed from an intermediate luminosity optical transient (ILOT) event. Clyne et al. (2014) further speculated on other nova-like formation events, some of which include planetary bodies that may be destroyed. Apart from the knots demonstrating similar mass and velocities to CN ejecta, we emphasise that these hypothetical CN or nova-like scenarios do not have any other observational support. Despite this, these scenarios appear to have been adopted as the de facto explanation for jet formation in MyCn~18 in the literature. As such, it is important to determine whether these hypothetical scenarios are physically compatible with the binary nucleus of MyCn~18. The observed orbital parameters of the binary can be assumed unchanged since the main nebula was ejected at the end of a CE interaction phase. These parameters uniquely enable us to determine whether any CN event could have produced the jet system since the main nebula was ejected. 

Classical novae occur in cataclysmic variables where the short orbital separation allows for mass transfer and accumulation onto the white dwarf leading to a thermonuclear explosion (Warner 1995). The time needed for the WD to build up a critical layer of hydrogen depends primarily on the mass of the WD and the accretion rate. While models for novae on degenerate WDs in thermal equilibrium are well developed (e.g. Yaron et al. 2005), the same cannot be said for pre-WDs that are found in PNe. Supposing accretion and thermonuclear explosions can occur on pre-WDs (which may also have strong winds), which is currently unclear as no models have explored the topic, then in the following we examine whether a CN could have launched the jets of MyCn~18. 

The strictest constraints come from the magnitude of permissible mass transfer rates. In CNe, the primary must have accreted a minimum amount of mass $M_\mathrm{env}\sim$10$^{-5}$--10$^{-4}$ $M_\odot$ for a 0.6 $M_\odot$ white dwarf (Yaron et al. 2005).\footnote{Shara et al. (2010) explored very rare luminous red novae that can occur on low-mass WDs ($M_\mathrm{WD}\lesssim0.65$ $M_\odot$) accreting at low rates ($\dot{M}_\mathrm{acc}\sim$10$^{-10}$--10$^{-12}$ $M_\odot$ yr$^{-1}$), however they must still accumulate $M_\mathrm{env}\sim10^{-4}$--10$^{-3}$ $M_\odot$ to ignite a nova explosion.} Considering that a CN launched jet must have accreted mass over the time span between the main nebula and jet formation (only $\sim$1050 yrs, Clyne et al. 2014), accreting the above $M_\mathrm{env}$ requires that the average accretion rate is an implausible $\dot{M}_\mathrm{acc}\sim$10$^{-8}$--10$^{-7}$ $M_\odot$ yr$^{-1}$ that cannot be supplied from the M5V companion. Even if the companion were an active M-dwarf, the accretion rate could only reach a maximum of $\dot{M}_\mathrm{acc}\sim10^{-11}$ $M_\odot$ yr$^{-1}$ during coronal mass ejections (Mullan 1996; Osten \& Wolk 2015), falling far short of the required rate. The inability to achieve sufficiently high accretion onto the primary from the companion therefore rules out any CN from powering the jets in MyCn~18.\footnote{Note that any accretion onto the primary would also be hindered by the strong wind of the primary which can reach $v_\infty\sim1000$ km s$^{-1}$ in Of-type central stars (Pauldrach et al. 2004).}

More generally, it is unlikely that CNe are in a position to produce jets in post-CE PNe. Jets are extremely rare in cataclysmic variables (K{\"o}rding et al. 2008, 2011) and are unlikely to account for the large numbers of post-CE PNe with jets. Furthermore, CNe are rarely observed to appear in PNe (e.g. Wesson et al. 2008) and the most promising additional candidates have been reclassified as old nova shells (e.g. Miszalski et al. 2016; Shara et al. 2017). In summary, while it is not completely unexpected for a CN to appear within a PN (e.g. Wesson et al. 2008), they are not expected to routinely form jets and are not common enough to play a significant role in producing jets of post-CE PNe. 

\subsection{A potential formation scenario for MyCn~18 involving fallback of CE ejecta} 
\label{sec:fallback}
The main hourglass nebula with its narrow waist was ejected about 2700 years ago (Clyne et al. 2014) likely at the end of a CE interaction phase. Preferential deposition of material in the orbital plane during the CE phase likely had a defining influence in shaping the bipolar nebula (e.g. Sandquist et al. 1998; Ricker \& Taam 2012; Passy et al. 2012). The inner hourglass nebula was ejected around 950 years later and was closely followed by the jet system 100 years later (Clyne et al. 2014). The close association in time between the inner hourglass and the jets strongly suggests they were formed from the same reservoir of matter. This matter would likely have been bound to the binary (circumbinary) or either component of the binary. As there could not have been any mass transfer between the binary components (Sect. \ref{sec:params}), the most likely source of this matter would be bound material left over from the CE phase (Passy et al. 2012; Ricker \& Taam 2012), which simulations suggest can fallback onto the binary or either component to form a disk of a few $\sim$0.1 $M_\odot$ (Kashi \& Soker 2011; Kuruwita et al. 2016). The observed jet velocities (O'Connor et al. 2000; Clyne et al. 2014) are consistent with being produced by this disk, whether the jets originate near the secondary\footnote{An M5 dwarf has an escape velocity of $\sim$582 km s$^{-1}$, assuming $M=0.20$ $M_\odot$ and $R=0.225$ $R_\odot$ (Bressan et al. 2012; Chen et al. 2014).}, the primary (Blackman \& Lucchini 2014), or around both stars. Another more speculative possibility is that a stellar or planetary tertiary companion provides the matter required (e.g. Clyne et al. 2014).\footnote{We emphasise that such scenarios were developed prior to knowledge of the binary system we have discovered in MyCn~18. Any inclusion of a tertiary stellar or planetary companion in formation scenarios concerning MyCn~18 must reproduce the observed binary system. As our observations do not allow us to determine whether a tertiary component is present, further discussion of the wide variety of such hypothetical scenarios is well beyond the scope of this paper.}

The self-similarity of the inner hourglass with the main nebula (Clyne et al. 2014) suggests the accreted matter may have formed a second CE (or mini-CE) that would likely lead to the formation of the inner hourglass and jets. Precisely how this process may have occurred in MyCn~18 remains to be determined with the aid of detailed simulations. The process may be similar to that outlined by Soker (2017) in which the formation of a circumbinary disk plays a central role and the jets facilitate the removal of any remaining envelope.

\subsection{The offset binary central star position}
\label{sec:offset}
The \emph{Hubble Space Telescope} imaging of MyCn~18 revealed the central star position to be offset from the geometric centre of each nebula component (Fig. \ref{fig:img}, Sahai et al. 1999; Clyne et al. 2014). The largest offset of 0.2'' occurs with respect to the inner hourglass and corresponds to $618\pm101$ au at our adopted distance of $3092\pm507$ pc. As with the high velocity jets, several scenarios have been hypothesised to explain how this offset was produced in the time between the main nebula was ejected and the jets were launched ($\sim$1050 yr, Clyne et al. 2014). Perhaps the most widely accepted explanation in the literature is that the offset is produced by asymmetric mass loss over long timescales on the AGB (Soker et al. 1998). This scenario appears unlikely to apply as the binary we have discovered has a current orbital separation of 0.124 au or 26.6 $R_\odot$, which is 80--800 times smaller than the expected final separation range of $\sim$10-100 au expected according to this scenario. If indeed the offset was produced during the AGB, we would expect a larger observed offset and especially a larger offset with respect to the main nebula.

An explosive CN or nova-like event causing a `kick' to the binary system to produce the offset has also been suggested by Sahai et al. (1999) and Clyne et al. (2014). As we have discussed (Sect. \ref{sec:nova}), the binary parameters do not allow for a CN or dwarf nova to form during this timeframe. One possibility may be that the material that formed the inner hourglass (see Sect. \ref{sec:fallback}) accreted directly onto the WD and produced a nova-like event.\footnote{It is unclear whether this can occur given the pre-WD nature of the primary which also exhibits strong winds.}

\begin{figure}
   \begin{center}
      \includegraphics[scale=0.4,bb=0 0 580 521]{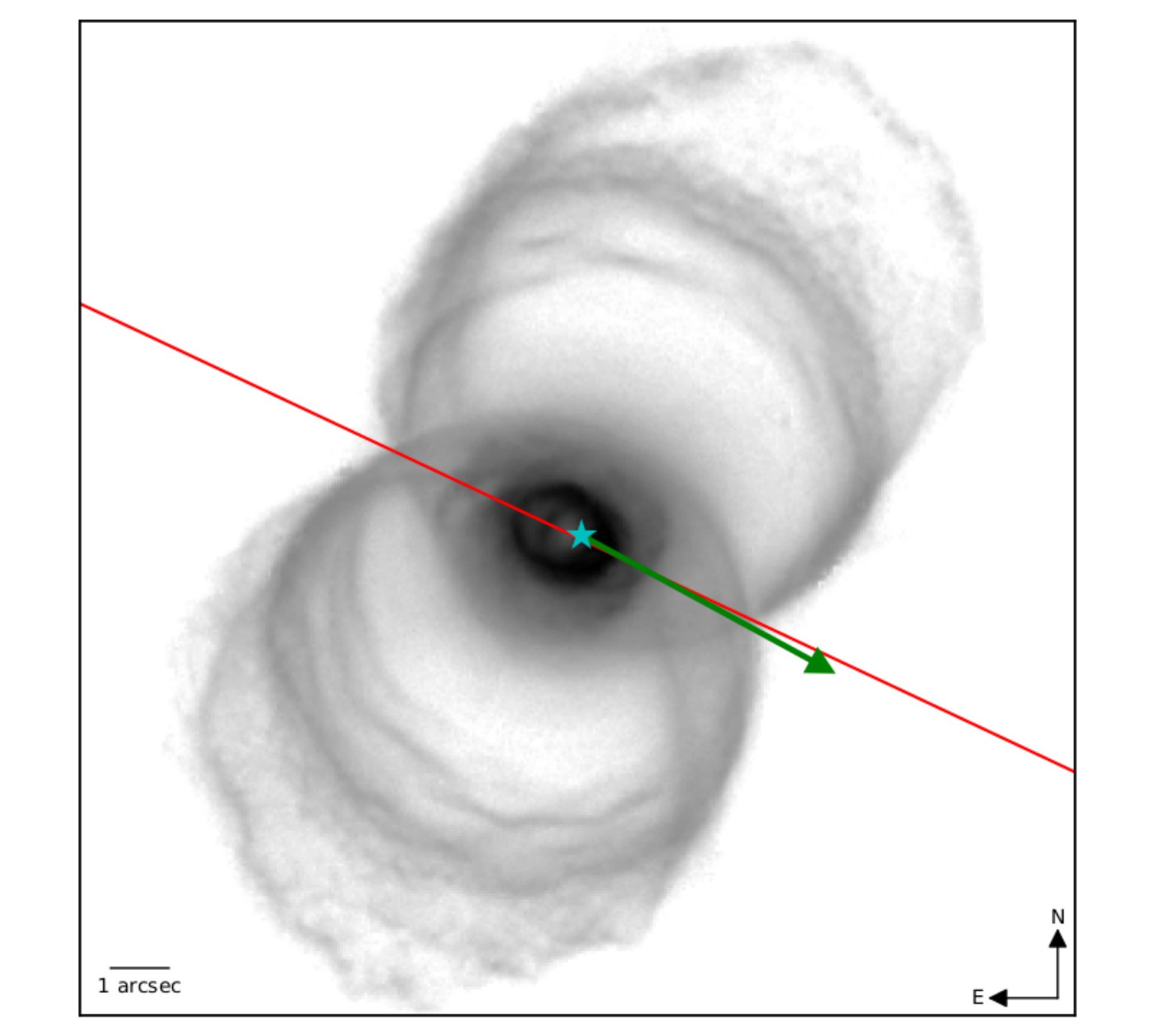}
   \end{center}
   \caption{\emph{Hubble Space Telescope} image of MyCn~18 taken with the $F656N$ filter showing the direction of proper motion (green arrow) of the central star (cyan star). The red line represents the minor axis.}
   \label{fig:proper}
\end{figure}

Perhaps the most promising explanation of the offset could be the proper motion. While Sahai et al. (1999) found this explanation problematic, we note that the direction of the offset matches the direction of the proper motion (Fig. \ref{fig:proper}) which is $\mu(\alpha)=-11.0\pm-1.1$ milliarcsec yr$^{-1}$ and $\mu(\delta)=-2.3\pm1.1$ milliarcsec yr$^{-1}$ (Zacharias et al. 2017). This match alone justifies a more thorough investigation than available data permit and is certainly beyond the scope of this paper. We suggest such a study would involve 3D smoothed particle hydrodynamics (SPH) simulations that include a more comprehensive prescription of the 3D geometry and properties of the inner region of MyCn~18\footnote{Clyne et al. (2014) considered only two slit positions covering the inner region. Additional slit positions would be beneficial to understand the elusive nature of the inner rings identified by Sahai et al. (1999).}  as well as higher quality proper motion data anticipated from the \emph{Gaia} mission.

\subsection{MyCn~18 and other post-CE PNe}

The 18.15 d orbital period of MyCn~18 lies between the 16 d orbit of NGC~2346 and the 142 d orbit of NGC~1360 (Miszalski et al. 2018). It is only the sixth binary central star known with a measured orbital period above 10 days (Miszalski et al. 2018). Population synthesis models predict large numbers of multiple day post-CE central stars (e.g. Nie et al. 2012), though it is unclear whether this population exists due to a historical tendency towards photometric surveys to find binary central stars (Miszalski et al. 2018). We will address this question empirically as our ongoing SALT HRS survey progresses, though we can say we have already discovered some new multiple day binaries. The multiple day systems appear to be very rare in the similar population of WD main-sequence binaries that has carefully studied selection effects (Nebot G{\'o}mez-Mor{\'a}n et al. 2011). If the CE phase operates similarly in central stars and WDMS binaries, then the initial SALT HRS discoveries might suggest multiple day post-CE binaries are more common than previously thought. Individually, multiple day post-CE central stars may be used to improve our understanding of CE population synthesis models (e.g. Davis et al. 2010) and simulations of the CE phase involving wider initial orbital separations (e.g. Iaconi et al. 2017).

Whether longer orbital periods favour the formation of bipolar PNe like NGC~2346 and MyCn~18 remains unclear and is currently biased by the small sample size of known binaries. The existence of canonical bipolar post-CE PNe with much shorter orbital periods, e.g. M~2-19 ($P=0.67$ d, Miszalski et al. 2009) and Hen~2-428 ($P=0.18$ d, Santander-Garc{\'{\i}}a et al. 2015), suggests there may be no simple correlation between orbital period and morphology. We also note that the often-cited formation scenario for bipolar PNe with narrow waists like MyCn~18 produces final orbital separations of $\sim$10--100 au (Soker \& Rappaport 2000), significantly larger than the observed 0.124 au separation of MyCn~18.

MyCn~18 is one of the few post-CE PNe with jets ejected after the CE phase. Jets ejected after the CE phase are less common and less understood than those ejected before the CE phase (Tocknell et al. 2014). At present the only other members include NGC~6337 (Garc{\'{\i}}a-D{\'{\i}}az et al. 2009; Hillwig et al. 2010) and NGC~6778 (Miszalski et al. 2011b; Guerrero \& Miranda 2012). Another poorly understood aspect of post-CE PNe are those exhibiting high abundance discrepancy factors (ADF) where abundances measured from optical recombination lines are higher than those derived from collisionally excited lines. Tsamis et al. (2004) measured a low ADF (O$^{2+}$) of 1.8 in MyCn~18. Sowicka et al. (2017) suggested that longer period systems tend to have low measured ADFs based on NGC~5189 (Manick et al. 2015; Garc{\'{\i}}a-Rojas et al. 2012) and IC~4776 (Sowicka et al. 2017). This depends on the purported 9 d orbital period of the central star of IC~4776 which is not definitive. The orbital period could be much shorter (Sowicka et al. 2017) and more observations are required before IC~4776 can be meaningfully compared against other post-CE PNe.

\section{Conclusions}
\label{sec:conclusion}
We presented SALT HRS \'echelle observations of the nucleus of the PN MyCn~18, also known as the Etched Hourglass Nebula. Radial velocity measurements from 26 spectra demonstrate a significant periodic variability of $18.15\pm0.04$ d. The data prove the presence of a post-CE binary nucleus in MyCn~18 which has long been suspected of being formed by a binary system. Several scenarios concerning the formation of MyCn~18 in the literature are not compatible with the orbital parameters of the binary nucleus. Our main conclusions are as follows:

\begin{itemize}
   \item The RV time series measured from the N~III $\lambda$4634 \AA\ stellar emission feature demonstrated a periodic variability of $18.15\pm0.04$ d at a significance level of $5\sigma$. The orbital period was used as the basis for a circular Keplerian orbit fit with a semi-amplitude of $11.0\pm0.3$ km s$^{-1}$ and a systemic velocity of $-71.70\pm0.24$ km s$^{-1}$. The latter is in good agreement with the nebula systemic velocity ($-71$ km s$^{-1}$, Clyne et al. 2014). Residuals of the Keplerian fit (2.61 km s$^{-1}$) are relatively high because of the influence of stellar winds in the Of(H) primary whose classification which we have clarified based on a deep stacked spectrum.
   \item Assuming a distance of $3092\pm507$ pc, we estimate the luminosity of the primary to be $\log L/L_\odot=3.76\pm0.31$. At $T_\mathrm{eff}=50\pm10$ kK, we estimate the primary mass to be $M_1=0.60\pm0.10$ $M_\odot$ and adopt an orbital inclination of $i=38\pm5$ deg determined from spatio-kinematic studies of the nebula (O'Connor et al. 2000; Clyne et al. 2014). The mass function derived from our observations then yields a secondary mass of $M_2=0.19\pm0.05$ $M_\odot$ corresponding to an M5 dwarf with an uncertainty of one spectral class.
   \item We rule out previous hypotheses that the jet system of MyCn~18 (O'Connor et al. 2000) formed as the result of a CN explosion. The orbital separation of 0.124 au or 26.6 $R_\odot$ is too large to allow for mass transfer via Roche-lobe overflow. Furthermore, an M5V secondary cannot provide the average accretion rate of $\dot{M}_\mathrm{acc}\sim10^{-8}$--10$^{-7}$ $M_\odot$ yr$^{-1}$ required to accrete the $M_\mathrm{env}\sim10^{-5}$--$10^{-4}$ $M_\odot$ mass required to form a CN (Yaron et al. 2005) in the $\sim1050$ yr timeframe between the main nebula ejection and when the jets are launched (Clyne et al. 2014).
   \item An alternative formation scenario for MyCn~18 was proposed whereby material from the CE ejection that formed the main nebula still bound to the binary settles or falls back to envelop the binary (e.g. Kashi \& Soker 2011; Kuruwita et al. 2016). This may have then lead to the formation of the inner hourglass and jets. The exact process in which this occurs remains uncertain and requires detailed simulations, though it could resemble that described by Soker (2017). There may be other mechanisms that contribute to forming MyCn~18, but these must be compatible with the observed binary system.
   \item We discussed the 0.2'' offset the central star shows with respect to the centre of the inner hourglass. The accepted scenario for producing observable offsets during the AGB (Soker et al. 1998) results in orbital separations 80--800 times larger than the observed separation in MyCn~18. The orbital parameters also rule out a CN or nova-like event to provide a `kick' to produce the observed offset (e.g. Sahai et al. 1999; Clyne et al. 2014). We favour the proper motion of MyCn~18 to explain the offset central star, especially since the offset is observed along the minor axis of MyCn~18 in the same direction as the proper motion. Detailed 3D SPH simulations of the motion of MyCn~18 that incorporates anticipated \emph{Gaia} proper motion data are encouraged to explore this possibility further.
   
   \item The discovery of the binary nucleus of MyCn~18 strengthens the long suspected link between bipolar nebulae and interacting binary stars (e.g. Soker \& Rappaport 2000). Characterising binary stars in bipolar nebulae across a wide variety of stellar masses and orbital separations may help shed light on common physical mechanisms behind the mass-loss histories of bipolar nebulae (e.g. Sugerman et al. 2005). One such common mechanism may be CE evolution, given the strong resemblance between MyCn~18 (a post-CE PN) and the nebular remnant of SN 1987A (a post-CE merger, Morris \& Podsiadlowski 2009). 

\end{itemize}

\begin{acknowledgements}
   This paper is based on spectroscopic observations made with the Southern African Large Telescope (SALT) under joint South African-Polish programmes 2016-2-SCI-034 and 2017-1-MLT-010 (PI: B. Miszalski). Polish participation in SALT is funded by grant No. MNiSW DIR/WK/2016/07. We are grateful to our SALT colleagues for maintaining the telescope facilities and conducting the observations. B. M. acknowledges support from the National Research Foundation (NRF) of South Africa. This study has been supported in part by the Polish MNiSW grant 0136/DIA/2014/43, and NCN grants DEC-2013/10/M/ST9/00086 and 2015/18/A/ST9/00746. H. V. W acknowledges support from the Research Council of the KU Leuven under grant number C14/17/082. We thank the anonymous referee for a helpful and constructive report. This paper also features observations made with the NASA/ESA Hubble Space Telescope, and obtained from the Hubble Legacy Archive, which is a collaboration between the Space Telescope Science Institute (STScI/NASA), the Space Telescope European Coordinating Facility (ST-ECF/ESA) and the Canadian Astronomy Data Centre (CADC/NRC/CSA). IRAF is distributed by the National Optical Astronomy Observatory, which is operated by the Association of Universities for Research in Astronomy (AURA) under a cooperative agreement with the National Science Foundation. B. M. thanks S. Mohamed for discussions and A. Y. Kniazev for making available his HRS pipeline data products. 
\end{acknowledgements}

\end{document}